\documentclass{elsart}

\usepackage{amssymb,amsmath,mathrsfs,eufrak}
\newtheorem{exm}{Example}[section]

\newcommand{\A}{\bf{A}}

\newcommand{\N}{\bf N}

\newcommand{\G}{\bf{G}}

\newcommand{\ar}{\rightarrow}
\newcommand{\la}[1]{\it {#1}\rm}

\newcommand{\T}{ {\bf T} }

\newcommand{\K}{ {\bf K} }

\newcommand{\1}{{\bf 1}}
\newcommand{\mN}{\mathbb{N}}

\newcommand{\Set}{ {\bf Set} }
\newcommand{\F}{\mathcal{F}}

\newcommand{\I}{{\bf I}}

\newcommand{\bS}{{\bf S }}

\newcommand{\df}[1]{\begin{defn} #1 \end{defn}}

\newcommand{\te}[1]{\begin{thm} #1 \end{thm}}
\newcommand{\ex}[1]{\begin{exm} #1 \end{exm}}

\newcommand{\rk}[1]{\begin{rem} #1 \end{rem}}

\newcommand{\X}{{\bf X }}

\newcommand{\TT}{{\mathsf T }}
\newcommand{\FF}{{\mathsf F }}

\begin{document}

\begin{frontmatter}



\title{Clones and Genoids  \\ in  \\ Lambda Calculus and First Order Logic }

%\large{(Research Announcement)}


{\large Zhaohua Luo}
\author{}

%\small{12/1/2007}

% \address{Geometry.net, 1800 W. Hillcrest 235, Newbury Park, CA, 91320}
 \ead{zluo@algebraic.net}
\ead[url]{http://www.algebraic.net/cag}

\begin{abstract}
% Text of abstract

A \la{genoid} is a category of two objects such that one is the
product of itself with the other. A genoid may be viewed as an
abstract substitution algebra. It is a remarkable fact that such a
simple concept can be applied to present a unified algebraic
approach to lambda calculus and first order logic.

\end{abstract}

%\begin{keyword}
%Clone \sep Universal Algebra \sep First-Order Theory \sep Lambda
%Calculus \sep Polyadic Algebra
%\PACS
%\end{keyword}
\end{frontmatter}

% main text

{\bf{Content \rm}}:
\\0. Introduction.
\\1. Genoids.
\\2. Clones.
\\3. Binding Algebras
\\4. Lambda Calculus.
\\5. First Order Logic
\\6. Clones Over A Subcategory.
\\7. Relate Work.

\section*{Introduction}
A \la{genoid} $(A, G)$ consists of a monoid $G$ with an element $+$,
a right act $A$ of $G$ with an element $x$, such that for any $a \in
A$ and $u \in G$ there is a unique element $[a, u] \in G$ such that
$x[a, u] = a$ and $+[a, u] = u$. A genoid represents a category with
two objects such that one is the dense product of itself with the
other.

Denote by $Act_G$ the category of right acts of $G$. The infinite
sequence of finite powers of $A$ in $Act_G$ determines a Lawvere
theory $Th(A, G)$:
\[A^0, A, \ A^2, \ A^3, \ ...\] A genoid may be viewed as a Lawvere theory with
extra capacity provided by $G$.

For any right act $P$ of $G$, we define a new right act $P^A = (P,
\circ)$, which has the same universe as $P$, but the action for any
$u \in G$ is defined by $a\circ u = a[x, u+]$. Let $ev: P^A \times A
\ar P$ be the map defined by $ev(p, a) = p[a, e]$ for any $p \in P$
and $a \in A$.  Then $\Lambda: hom(T \times A, P) \ar hom(T, P^A)$
defined by $(\Lambda f)t = f(t+, x)$ for any $t \in T$ is bijective,
with the inverse $\Lambda': hom(T, P^A) \ar  hom(T \times A, P)$
defined by $(\Lambda'g)(t, a) = ev(g(t), a)$. Thus $(P^A, ev)$ is
the exponent in the cartesian closed category $Act_G$. In particular
if $T = P = A$ we obtain a canonical bijection \[ \Lambda: hom(A
\times A, A) \ar hom(A, A^A).\]  This is the starting point of
lambda calculus.

We define an \la{extensive lambda genoid} to be a genoid $(A, G)$
together with two homomorphisms $\lambda: A^A \ar A$ and $\bullet: A
\times A \ar A$ such that $(\Lambda \bullet) \lambda = id_A$
($\beta$-conversion) and $\lambda (\Lambda \bullet) = id_A$
($\eta$-conversion). This means that $A$ and $A^A$ are isomorphic as
right acts of $G$. Conversely, any genoid $(A, G)$ such that $A$ and
$A^A$ are isomorphic determines an extensive lambda genoid.

A \la{quantifier algebra of a genoid $(A, G)$} is a Boolean algebra
$P$ which is also a right act of $G$ with Boolean algebra
endomorphisms as actions, together with a homomorphism $\exists: P^A
\ar P$ such that $\exists (p\vee q) = (\exists p) \vee (\exists q)$, $ p \leq (\exists p)+$, and $(\exists p+) = p$ for any $p, q \in P$. The study of a first
order theory can also be reduced to the study of a quantifier
algebra for a genoid $(A, G)$.

We say a genoid $(A, G)$ is a \la{clone} if $G$ is the countable
power $A^{\omega}$ of $A$. Algebraically the class of clones forms a
(non-finitary) variety. A general theory of \la{clones over any full
subcategory of a category} is presented at the end of this paper.

The theory of clones considered in this paper originated from the
theory of monads. Two equivalent definitions of monads, namely
\la{monads in clone form} and \la{monads in extension form} given by
E. Mane~\cite{mane}, can be interpreted as only defined over a given
subcategory of a category. These are \la{clones in algebraic form}
and \la{clones in extension form} over a subcategory respectively.
It turns out that these two forms of clones are no longer equivalent
unless the subcategory is dense. But morphisms of clones, algebras
of clones, and morphisms of algebras can all be defined for these
two types of clones. Since many familiar algebraic structures, such
as monoids, unitary Menger algebras, Lawvere theories, countable
Lawvere theories, classical and abstract clones are all special
cases of clones over various dense subcategories of $\Set$, the
syntax and semantics of these algebraic structures can be developed
in a unified way, so that it is much easier to extend these results
to many-sorted sets.

%The reader is referred to ~\cite{sa:1} For the related

\section{Genoids}

A \la{genoid theory} is a category $(\A, \G)$ of two objects
together with two morphisms $x: \G$$ \ar \A$ and $+: \G$$ \ar \G$
such that $(\G$$, x, +)$  is the product of $\A$ and $\G$, i.e. $\G
= \A \times \G$. We also assume that $\G$ is a dense object,
although this is not essential.  A \la{left algebra} of $(\A, \G)$
is a functor from $(\A, \G)$  to the category $\Set$ of sets
preserving the product. A \la{morphism of left algebras} is a
natural transformation.

Recursively we have
\[{\bf G \rm} = {\bf A \rm}^n \times \G = A \times ... \times A \times \G\]
for any positive integer $n$. If $\G$ is the countable power of
$\A$, i.e.
\[\G = \A^{\omega} = \A \times \A \times ...\] then we say that
$(\A, \G)$ is a \la{clone theory}.

Let $A = hom(\G, \A)$ and $G = hom(\G, \G)$. Then $G$ is a monoid
and $A$ is a right act of $G$. For any pair $(a, u) \in A \times G$
let $[a, u]: \G \ar \G$ be the unique morphism such that $x[a, u] =
a$ and $+[a, u] = u$. Then $u = [xu, +u]$ for any $u \in G$.

\df{A genoid $(A, G, x, +, [ \ ])$ consisting of a monoid $(G, e)$,
a right act $A$ of $G$, $x \in A$, $+ \in G$, and a map $[ \ ]: A
\times G \ar G$ such that for any $a \in A$ and $u \in G$ we have

(G1) $x[a, u] = a$.

(G2) $+[a, u] = u$.

(G3) $u =  [xu, +u]$.}

A genoid is simply denoted by $(A, G)$. Clearly any genoid theory
$(\A, \G)$ determines a genoid $(A, G)$. Since we assume $\G$ is a
dense object, $(\A, \G)$ is uniquely determines by $(A, G)$.
Conversely if $(A, G)$ is a genoid then the subcategory of right
acts of $G$ generated by $A$ and $G$ is a genoid theory. Hence the
notions of genoid theory and genoid are equivalent.

\rk{Genoids form a variety of $2$-sorted heterogeneous finitary
algebras with universes $A$ and $G$. A genoid $(A, G)$ is called
standard if it is generated by $A$ as a 2-sorted algebra. }

Suppose $(A, G)$ is a genoid. We have $e = [x, +]$ and $[a, u]v =
[av, uv]$ for any $a \in A$ and $u, v \in G$. Thus $+ = e+ = [x, +]+ = [x+, ++]$. We shall write $[a_1,
a_2, ..a_n, u]$ for $[a_1, [a_2, [... [a_n, u]...]]]$.

Let $x_1 = x$, $x_{i+1} = x_i+$  for any $i > 0$. Then axiom (G3)
extends to
\[u  = eu = [x, +]u = [x_1u, +u] = [x_1u, [x+, ++]u] \]\[ = [x_1u, [x_2u, +^2u]] = ...  = [x_1u, x_2u, ... , x_nu, +^nu]\] for any $n > 0$.

It is easy to define a \la{many-sorted genoid} for a nonempty set
$S$ of sorts:

\df{An $S$-genoid theory is a category $(\{\A$$^s\}_{s \in S},$ $
\G)$ of objects $\{\A$$^s\}_{s \in S}$ and $\G$, together with
morphisms $\{x^s: \G$ $ \ar \A$$^s\}_{s\in S}$ and $\{+^s: \G$$ \ar
\G$$\}_{s\in S}$ such that $(\G$$, x^s, +^s)$ is the product of
$\A^s$ and $\G$ for all $s \in S$, and $x^s+^t = x^s$, $+^s+^t =
+^t+^s$ for any distinct $s, t \in S$. We also assume that $\G$ is a
dense object. A left algebra of $(\{\A$$^s\}_{s \in S},$ $ \G)$ is a
functor from this category to $\Set$  preserving the products. }

\df{An $S$-genoid is a pair $(A, G)$ consisting of a monoid $G$ and
a set $A = \{(A^s, G, x^s, +^s, [ \ ]^s)\}_{s \in S}$ of genoids
such that $x^s+^t = x^s$  and $+^s+^t = +^t+^s$ for any two distinct
elements $s, t \in S$.}

Suppose $(A, G)$ is an $S$-genoid. For any $s \in S$ let
$\kappa^s_n: G \ar (A^s)^n$ be the map sending each $u \in G$ to
$[x^s_1u, x^s_2u, ..., x_n^su] \in (A^s)^n$. Let $(T, n)$ be a pair
consisting of a finite subset $T$ of $S$ and an integer $n > 0$. We
say an element $p$ of $P$ \la{has a finite support $(T, n)$ (or $p$
has a finite rank $n$)} if $pu = pv$ for any $u, v \in G$ with
$\kappa^s_n(u) = \kappa^s_n(v)$ for any $s \in T$. We say $p$
\la{has a finite rank $0$} (or \la{$p$ is closed}) if $pu = pv$ for
any $u, v \in G$. An element of $P$ is called \la{finite} if it
has a finite support. We say $P$ is \la{locally finite} if any of
its element is finite. We say $(A, G)$ is a \la{locally finite
genoid} if $A$ is locally finite as a right act of $G$.

\ex{An  algebraic genoid  is a monoid $G$ together with two elements
$x, + \in G$ such that $xx = +x = x$, and $(xG, G, x, +)$ is a
genoid. Algebraic genoids form a finitary variety.}

\ex{An algebraic $S$-genoid with a zero element $0$ is a monoid $G$
with a zero element $0$ together with a set $\{x^s, +^s\}_{s\in S}$
of pairs of elements of $G$ such that

1. $x^sx^t = 0$ for any distinct $s, t \in S$.

2. $+^sx^t = x^tx^t = x^t$ for any $s, t \in S$.

3. $(A, G)$ is an $S$-genoid with $A = \{(x^sG, G, x^s, +^s)\}_{s\in
S}$.

Algebraic $S$-genoids form a finitary variety. }

\section{Clones}

Let $\mN$ be the set of positive integers.

\df{A clone theory over $\mN$ is a category $(\A, \G)$ of two
objects together with an infinite sequence of morphisms $x_1, x_2,
..$ from $\G$ to $\A$ such that $(\G$, $\{x_1, x_2, ..\})$ is a
countable power of $\A$. A left algebra of $(\A, \G)$ is a functor
from $(\A, \G)$ to $\Set$ preserving the countable power.}

Let $A = hom(\G, \A)$ and $G = hom(\G, \G)$. Then $G$ is a monoid
and $A$ is a right act of $G$. For any infinite sequence $a_1, a_2,
....$ of elements of $A$ let $[a_1, a_2...] \in G$ be the unique
morphism such that $x_i[a_1, a_2...] = a_i$ for any integer $i > 0$.
Then $u = [x_1u, x_2u,...]$ for any $u \in G$. Any clone theory
determines a genoid theory with $x = x_1$, and $+ = [x_2, x_3,
...]$.

\df{A clone in extension form over $\mN$ is a nonempty set $A$ such
that

(i) The set $A^*$ of all the infinite sequences $[a_1, a_2, ...]$ of
elements of $A$ is a monoid with a unit $[x_1, x_2, ...]$.

(ii) $A$ is a right act of $A^*$.

(iii) $x_i[a_1, a_2, ...] = a_i$ for any $[a_1, a_2, ...]$ and $i >
0$.}

Any clone $A$ in extension form determines a genoid $(A, A^*, x_1,
+, [ \ ])$ with $+ = [x_2, x_3, ...]$ and $[a_1, [b_1, b_2 ,..]] =
[a_1, b_1, b_2 ,..]$. Thus we may speak of locally finite clones.
Conversely, assume $(A, G)$ is a any genoid. Denote by $\F(A)$ the
set of finite elements of $A$. For any $a \in \F(A)$ with a finite
rank $n > 0$ and $[a_1, a_2, ...] \in \F(A)^*$ define
\[a[a_1, a_2, ...] = a[a_1, a_2, ..., a_n , e],\] which is independent of
the choice of $n$. Let \[[a_1, a_2, ...][b_1, b_2, ...] = [a_1[b_1,
b_2, ...], a_2[b_1, b_2, ...], ...].\] Then $\F(A)^*$ is a monoid
with the unit $[x_1, x_2, ...]$, $\F(A)$ is a right act of $A^*$,
and $x_i[a_1, a_2, ...] = a_i$. Thus $\F(A)$ is a locally finite
clone. If $A = \F(A)$ is locally finite then we have a canonical
homomorphism of genoids $(A, G) \ar (A, A^*)$ sending each $u \in G$
to $[x_1u, x_2u, ... x_nu, ...]$.

\rk{Clones form a variety of infinitary algebras with universe $A$.
}

\rk{The notion of a locally finite clone is equivalent to that of
a Lawvere theory (without the $0$-ary object). }

\df{Let $A$ be a clone. A left algebra of $A$ (or a left
$A$-algebra) is a set $D$ together with a multiplication $A \times
D^{\mN} \ar D$ such that for any $a \in A$, $[a_1, a_2, ...] \in
A^{\mN}$ and $[d_1, d_2, ...] \in D^{\mN}$ \\1. $(a[a_1, a_2,
...])[d_1, d_2, ...] = a([a_1[d_1, d_2, ...], a_2[d_1, d_2, ...],
...]$. \\2. $x_i[d_1, d_2, ...] = d_i$.}

\rk{Left algebras of clones are main objects of study in universal
algebra (cf.~\cite{luo:2}).}

\df{A clone in algebraic form over $\mN$ is a nonempty set $A$ such
that the set $A^{\mN}$ of maps from $\mN$ to $A$ is a monoid and
$(ru)v = r(uv)$ for any maps $r: N \ar N$ and $u, v: N \ar A$. }

\rk{Since $\mN$ is dense in $\Set$, one can show easily that the two
forms of clones over $\mN$ are equivalent. Therefore in the
following we shall not distinguish these two types of clones.}

\section{Binding Algebras}\label{sec:ba}

Let $(A, G)$ be a genoid. The map  $\delta: G \ar G$ sending $u$ to
$[x, u+]$ is an endomorphism of monoid $G$. Let $- = [x, e]$. One
can show that $(\delta, +, -)$ is a monad on the one-object category
determined by the monoid $G$, as we have $+- = (\delta+)- = e$ and
$-- = (\delta-)-$. The Kleisli category of this monad is the monoid
$(G, *)$ with $u*v = u[x_1, v]$.

If $P$ is any right act of $G$ denote by $P^A$ the new right act
$(P, \circ)$ of $G$ defined by $p\circ u = p(\delta u) = p[x, u+]$
for any $p \in P$ and $u \in G$. Let $ev: P^A \times A \ar P$ be the
map defined by $ev(p, a) = p[a, e]$. Define $\Lambda: hom(T \times
A, P) \ar hom(T, P^A)$ by $(\Lambda f)t = f(t+, x)$ for any $t \in
T$, and $\Lambda': hom(T, P^A) \ar  hom(T \times A, P)$ by
$(\Lambda'g)(t, a) = ev(g(t), a)$. Then both $\Lambda' \Lambda$ and
$\Lambda \Lambda'$ are identities (which implies that $\Lambda$ is
bijective). Thus $(P^A, ev)$ is the exponent in the the cartesian
closed category $Act_G$ of right acts of $G$.

Let $\Delta: Act_G \ar  Act_G$ be the functor sending each act $P$
to $P^A$, and each morphism $f: P \ar Q$ to $f: P^A \ar  Q^A$. The
actions of $+$ and $-$ induces two natural transformations $+ : Id
\ar \Delta$ and $-: \Delta^2 \ar \Delta$. It is easy to see that
$(\Delta, +, -)$ is a monad on $Act_G$.

\df{A binding operation is a homomorphism $P^A \ar P$. A cobinding
operation is a homomorphism $P \ar P^A$. }

\rk{We assume $y, z, w, ..., y_1, y_2, ... , z_1, z_2, ... \in
\{x_1, x_2 , x_3 ...\}$, which are called syntactical variables.
Suppose $\sigma$ is a binding operation. The traditional operation
$\sigma x_i : P \ar P$ (for each variable $x_i$) is defined as the
derived operation:
\[\sigma x_i.p = \sigma (p[x_2, x_3, ..., x_i, x_1, +^{i+1}]).\] If
$y = x_i$ then $\sigma y.p$ means $\sigma x_i.p$. }

\ex{For any $p \in P$ we have

1. $\sigma x_1.p = \sigma (p[x_1, ++]) = (\sigma p)+$.

2. $\sigma p = (\sigma x_1.p)-$.

3. If $p$ has a finite rank $n > 0$ then $\sigma p$ has a finite
rank $n - 1$. Thus $\sigma^n p$ is closed.

4. If $p$ is closed then $\sigma p$ is closed. }

\df{Let $S$ be any set of sorts. Let $k$ be a non-negative integer.
An $S$-arity of rank $k$ is a finite sequence $\alpha = <(s_1,
n_1),..., (s_k, n_k), (s_{k+1}, n_{k+1})>$ with $s_i \in S$ and $n_i
\geq 0$. An $\alpha$-binding operation on a right act $P$ of an
$S$-genoid $(A, G)$ is a homomorphism of right acts
\[(\Delta^{s_1})^{n_1}
 P \times ... \times
 (\Delta^{s_k})^{n_k} P \ar
(\Delta^{s_{k+1}})^{n_{k+1}}P\] (assume $(\Delta^{s_i})^0 P = P$),
where $\Delta^{s_i}$ is the functor sending each right act $Q$ of
$G$ to $Q^{A^{s_i}}$. }

\df{ An $S$-signature  is a set $\Sigma$ of operation symbols such
that for each symbol $f \in \Sigma$ an $S$-arity $ar(f)$ is
attached. A $\Sigma$-binding algebra for an $S$-genoid $(G, A)$ is a
right act $P$ of $G$ such that for each symbol $f \in \Sigma$ an
$ar(f)$-binding operation $f^P$ on $P$ is assigned. }

\rk{We shall drop all the references to the elements of $S$ if $S$
is a singleton.  Thus an arity of rank $k$ is simply a finite
sequence $\alpha = <n_1,..., n_k, n_{k+1}>$ of non-negative
integers. }

 \ex{1. A binding operation is a $<1,
0>$-operation.

2. A cobinding is a $<0, 1>$-operation.

3. A homomorphism $P^2 \ar P$ is a $<0, 0, 0>$-operation.

4. A homomorphism $P^0 \ar P$ is a $<0>$-operation, which reduces to
a closed element of $P$.}

Lambda genoids and predicate algebras defined below are examples of
binding algebras.

\section{Lambda Calculus}

A genoid $(A, G)$  is \la{reflexive} if $A^A$ is a retract of $A$
(as right acts of $G$).  It is \la{extensive} if $A^A$ is isomorphic
to $A$.

A \la{lambda genoid} is a genoid $(A, G)$ together with two
homomorphisms $\lambda: A^A \ar A$ and $\cdot: A^2 \ar A$ of right acts of
$G$. If $((\lambda a)+) x = a$ for any $a \in A$  we say $A$ is a
\la{reflexive lambda genoid} (or a $\lambda_{\beta}$-genoid). If
furthermore $\lambda((a+) x) = a$ for any $a \in A$  then we say $A$
is an \la{extensive lambda genoid} (or a $\lambda_{\beta
\eta}$-genoid). Thus a genoid is reflexive (resp. extensive)  iff it
is the underlying genoid of a reflexive (resp. extensive) lambda
genoid.

\rk{Lambda clones (resp. reflective lambda clones, resp. extensive
lambda clones) form a variety of (infintary) algebras. The initial
lambda clone is precisely the clone determined by terms in $\lambda
\sigma$-calculus (cf~\cite{ha:1}).}

\rk{Lambda algebraic genoids (resp. reflective lambda algebraic
genoids, resp. extensive lambda algebraic genoids) form a variety of
finitary algebras.}

\rk{The classical operation $\lambda x_i : A \ar A$ (for each
variable $x_i$) is defined as the derived operation:
\[\lambda x_i.a = \lambda (a[x_2, x_3, ..., x_i, x_1, +^{i+1}]).\] If
$y = x_i$ then $\lambda y.a$ means $\lambda x_i.a$.}

Suppose $(A, G)$ is an extensive lambda genoid. Assume $a, b, c \in
A$ and $u \in G$. Here are some useful formulas:

(1) $(\lambda a)b = a[b, e]$.

(2) $((\lambda a)u)b = a[b, u]$.

(3) $(\lambda a+)b = a$.

(4) $(\lambda^n a)+^nx_n...x_1 = a$ for any integer $n > 0$.

(5) If $a$ has a finite rank $n > 0$ then $\lambda^na$ is closed and
$(\lambda^na)x_n...x_1 = a$. Thus
\[(\lambda^na)a_n...a_1 = (\lambda^na)x_n...x_1 [a_1, ..., a_n, e]  = a[a_1, ...,
a_n]\] (6) An element $a$ has a finite rank $n > 0$ if and only if
there is a closed element $c$ such that
\[a = cx_n...x_1.\]
The following closed terms play important roles in lambda calculus
(notation: $\lambda y_1...y_n.a = \lambda y_1.(\lambda
y_2.(..(\lambda y_n.a)...))$.)

$\I = \lambda y.y =  \lambda x_1$.

$\K = \lambda yz.y = \lambda \lambda x_2$.

$\bS  = \lambda yzw.yw(zw) = \lambda \lambda \lambda x_3x_1(x_2
x_1)$.

It follows from (5) we have

$\I a =  x_1 [a, e] = a$.

$\K ab = x_2[b, a, e] = a$.

$\bS abc = (x_3x_1(x_2x_1))[c, b, a, e] = ac(bc)$.

\df{Let $S$ be a nonempty set carrying a binary operation $\ar$. An
$S$-simply typed lambda genoid  is an $S$-genoid $(A, G)$ together
with homomorphisms $\{\lambda^s: (A^t)^{A^s} \ar A^{s\ar t}\}$ and
$\{A^{s \ar t} \times A^s \ar  A^t\}$ such that for any $a \in A^t$
and $c \in A^{s\ar t}$ we have $(\lambda^s a)+^sx^s = a$ and
$\lambda^s(c+^sx^s) = c$.}

\section{First Order Logic}\label{sec:fol}

A \la{predicate algebra of an $S$-genoid} $(A, G)$ is a right act
$P$ of $G$ together with homomorphisms of right acts $\{\exists^s:
P^{A^s} \ar P\}_{s \in S }$, $\FF: P^0 \ar P$, and $\Rightarrow: P
\times P \ar P$.

Define the following derived operations on $P$:

$\neg p = (p \Rightarrow \FF)$,

$\TT = \neg \FF$,

$p \vee q = (\neg p) \Rightarrow q$,

$p \wedge q = \neg (p \Rightarrow \neg q)$.

We say  $P$ is a \la{reduced predicate algebra} if for any $p,q, \in
P$ and $s \in S$

(i) $(\vee, \wedge, \neg, \FF, \T)$ defines a Boolean algebra $P$.

(ii) $\exists^s (p\vee q) = (\exists^s p) \vee (\exists^s q)$.

(iii) $ p \leq (\exists^s p)+^s$.

(iv) $ \exists^s (p+^s) = p$.

A reduced predicate algebra is also called a \la{quantifier
algebra}.

\rk{The class of predicate algebras (resp. quantifier algebras) of a
genoid is a variety of finitary algebras.}

An \la{interpretation of a predicate algebra $P$} is a pair $(Q,
\mu)$ consisting of a reduced predicate algebra $Q$ and a
homomorphism $\mu: P \ar Q$ of predicate algebras. We say $p \in P$
is \la{logical valid} (written $\models p$) if for any
interpretation $(Q, \mu)$ we have $\mu(p) = \TT$. If $p, q \in P$
then we say that $p$ and $q$ are \la{logically equivalent} (written
$p \equiv q$) if $(p \Rightarrow q) \wedge (q \Rightarrow p)$ is
logically valid. Then $\equiv$ is a congruence on $P$. The set of
congruence classes of $P$ with respect to the congruence $\equiv$ is
a reduced predicate algebra called the \la{Lindenbaum-Tarski algebra
of $P$}. (see~\cite{luo:2} for a further development of the theory
of predicate algebras).

\te{ Suppose $A$ is a locally finite clone. Any left algebra $D$
of $A$ determines a locally finite reduced predicate algebra
$P(D^{\mN})$, where $P(D^{\mN})$ is the power set of $D^{\mN}$. A
locally finite predicate algebra of $A$ is reduced iff it belongs
to the variety generated by predicate algebras $P(D^{\mN})$ for all
left algebras $D$ of $A$. }

\section{Clones Over A Subcategory}

\df{Let $\N$ be a full subcategory of a category $\X$. A clone (in
extension form, or Kleisli triple) over $\N$ is a system $T = (T,
\eta, \ ^*-)$ consisting of functions

(a) $T: Ob \N$$  \ar  Ob \X$.

(b) $\eta$ assigns to each object $A \in \N$ a morphism $\eta_A: A
\ar TA$.

(c) $*-$ maps each morphism $f: B \ar TC$ with $B, C \in \N$ to a
morphism $*f: TB \ar TC$, such that for any $g: C \ar TD$ with $D
\in \N$

(i) $*f*g = *(f*g )$.

(ii) $\eta_B*f = f$.

(iii) $*\eta_C = id_{TC}$. }

\rk{If $\N = \X$ we obtain the original definition for a Kleisli
triple over a category, which is an alternative description of a
monad.}

\df{Let $\N$ be a full subcategory of a category $\X$. A clone
theory in extension form (resp. in algebraic form) over $\N$ is a
pair $(\K, T)$ where $\K$ is a category and $T$ is a functor $T: \K
\ar \X$ (resp. $T$ is a function $T: Ob \K \ar Ob \X$) such that for
any $A, B, C, D \in \N$

(i) $Ob \N$$ = Ob \K$.

(ii) $\K(A, B) = \X(A, TB)$.

(iii) $f(Tg) = fg$ (resp. $r(fg) = (rf)g$) for any $f \in \K(A, B)$,
$g \in \K(B, C)$ and $r \in \N$$(D, A)$.}

\rk{If $\N$ is dense then these two forms of clone theory are
equivalent. In particular, a clone theory over  $\N = \X$ in both
forms corresponds to a monad on $\X$.}

\rk{Any clone over $\N$ determines a clone theory in extension form
over $\N$, called its Kleisli category. Conversely any clone theory
in extension form over $\N$ induces a clone over $\N$
(see~\cite{luo:2} for details).}

\ex{ Let $\X = \Set$ be the category of sets.

1. A clone over a singleton is equivalent to a monoid.

2. A clone over a finite set is equivalent to a unitary  Menger
algebra.

3. A clone over a countable set is equivalent to a clone over $\mN$
defined above.

4. A clone over the subcategory $\{0, 1, 2, ...\}$ of finite sets is
equivalent to a clone in the classical sense (i.e. a Lawvere
theory).}

\rk{1. A clone theory over $\N = \X$  is equivalent to a monad on
$\X$.

2. A clone (or a monad) over a one-object category is called a
Kleisli algebra.

3. Any genoid $(A, G)$ determines a Kleisli algebra since $(\delta,
+, -)$ is a monad on the one-object category $G$.}

\section{Relate Work}

In classical universal algebra one studies left algebras of a clone
over $\mN$. Such a clone can be defined in many different ways
(see~\cite{cohn:1}\cite{newmann:rep}\cite{parho:llinear}\cite{schtro:algmul}).
Our approach to binding algebras and untyped lambda calculus was
greatly inspired by~\cite{fi:1}. For other algebraic approaches to
untyped lambda calculus
see~\cite{ha:1}\cite{ja:1}\cite{ma:1}\cite{ob:1}\cite{sa:1}\cite{se:1}.
Our definition of a quantifier algebra of a genoid was based on
Pinter~\cite{pi:1} (see also~\cite{ci:1}\cite{pl:1}).

\end{document}